\documentclass[prl,twocolumn, aps,superscriptaddress,floatfix,nofootinbib]{revtex4}

\usepackage{graphicx}
\usepackage{verbatim}
\usepackage{mathrsfs}
\usepackage{amsmath,amsfonts,amssymb}
\usepackage{graphicx}

\def\3{2.8in}    
\def\2{2.5in}
\def\4{3.0in}

\def \beq {\begin{equation}}
\def \eeq {\end{equation}}
\begin{document}

\title{Search for superconducting proximity effect in a topological insulator and high temperature superconductor heterostructure Bi$_2$Se$_3$/Bi$_2$Sr$_2$CaCu$_2$O$_{8+\delta}$}

\author{Su-Yang~Xu}
\affiliation {Joseph Henry Laboratory, Department of Physics, Princeton University, Princeton, New Jersey 08544, USA}

\author{Chang~Liu}
\affiliation {Joseph Henry Laboratory, Department of Physics, Princeton University, Princeton, New Jersey 08544, USA}

\author{Anthony Richardella}
\affiliation {Department of Physics, The Pennsylvania State University, University Park, Pennsylvania 16802-6300, USA}

\author{I.~Belopolski}
\affiliation {Joseph Henry Laboratory, Department of Physics, Princeton University, Princeton, New Jersey 08544, USA}

\author{N.~Alidoust}
\affiliation {Joseph Henry Laboratory, Department of Physics, Princeton University, Princeton, New Jersey 08544, USA}

\author{M.~Neupane}
\affiliation {Joseph Henry Laboratory, Department of Physics, Princeton University, Princeton, New Jersey 08544, USA}

\author{G. ~Bian}
\affiliation {Joseph Henry Laboratory, Department of Physics, Princeton University, Princeton, New Jersey 08544, USA}


\author{Nitin Samarth}
\affiliation {Department of Physics, The Pennsylvania State University, University Park, Pennsylvania 16802-6300, USA}

\author{M.~Z.~Hasan}
\affiliation {Joseph Henry Laboratory, Department of Physics, Princeton University, Princeton, New Jersey 08544, USA}
\affiliation {Princeton Center for Complex Materials, Princeton University, Princeton, New Jersey 08544, USA}


\date{\today}

\begin{abstract}
We probe the near Fermi level electronic structure of tunable topological insulator (Bi$_2$Se$_3$)-cuprate superconductor Bi$_2$Sr$_2$CaCu$_2$O$_{8+\delta}$  ($T_{\textrm{c}}\simeq91$ K) heterostructures in their proximity induced superconductivity regime. In contrast to previous studies, our careful momentum space imaging provides clear evidence for a two-phase coexistence and a lack of $d$-like proximity effect. Our Fermi surface imaging data identifies major contributors in reducing the proximity-induced gap below the 5 meV range. These results correlate with our observation of momentum space separation between the Bi$_2$Se$_3$ and Bi$_2$Sr$_2$CaCu$_2$O$_{8+\delta}$ Fermi surfaces and mismatch of crystalline symmetries in the presence of a small superconducting coherence length.  These studies not only provide critical momentum space insights into the Bi$_2$Se$_3$/Bi$_2$Sr$_2$CaCu$_2$O$_{8+\delta}$ heterostructures, but also set an upper bound on the proximity induced gap for realizing much sought out Majorana fermion condition in this system.
\end{abstract}


\maketitle

Topological order proximity to superconductivity (SC) is of central interest in condensed matter physics \cite{Kane_Proximity, Zhang_TSC, Patrick Lee, Linder, Sarma, SUSY1, RMP, Zhang_RMP}. A wide range of topological quantum phenomena such as $p+ip$-wave topological superconductivity, non-Abelian zero modes, and supersymmetry physics have been theoretically predicted, if superconductivity can be induced in the helical topological surface states of a 3D topological insulator (TI) \cite{Kane_Proximity, Zhang_TSC, Patrick Lee, Sarma, SUSY1, Linder}. Following the theoretical proposals, many transport and scanning tunneling microscopy (STM) experiments have been performed to study the superconducting proximity effect between TI materials and superconductors \cite{Hor, Ando, Nitin, Kanigel, Morpurgo, LuLi1, Mason, LuLi2, Gordon, Dong, Brinkman, Burch, Molenkamp, Kirzhner, Millo, LuLi3, Leo, Shtrikman}. Important progress has been reported in this context, including the demonstration of super-current \cite{Morpurgo, Mason} and superconducting gap \cite{Dong} in TI materials and the observation of zero-bias transport anomalies \cite{Ando, Leo, Shtrikman}. However, a concrete and decisive observation of superconductivity in the helical surface states remains lacking, limiting the unambiguous realization of topological superconductivity and Majorana fermions. Demonstration of superconductivity in the helical surface states is challenging due to the coexistence of helical surface states, bulk bands, impurity states and possible trivial surface states at the Fermi level, separating all these contributions using transport and STM probes is limited by there inherent lack of the momentum and spin resolution. Such demonstration is further important since recent theoretical and experimental efforts have shown that the coexistence of irrelevant electronic states at the Fermi level can lead to ambiguous interpretations of transport and STM signals \cite{Patrick Lee2, TeWari, Marcus, Franceschi}. Therefore, a concrete and decisive observation of superconductivity in the helical surface states not only serves as a cornerstone for realizing the proposed exotic new physics, but also can be used as an experimental methodology for developing a material-sample feedback loop in order to measure, isolate and enhance the ratio of Cooper pairing in the surface states from that of the irrelevant and undesirable Cooper pairing from the bulk, trivial, and impurity states.

A recent photoemission experiment has drawn particular attention because it reported the observation of a large gap ($\simeq15$ meV) in the electron-quasiparticle density of states (interpreted as the superconducting gap) in the surface states of TI Bi$_2$Se$_3$ films (even as thick as $7$ quintuple layer $\simeq7$ nm) grown on top of a $d$-wave high temperature superconductor Bi$_2$Sr$_2$CaCu$_2$O$_{8+\delta}$ (BSCCO) \cite{Zhou}. However, the gap in Ref. \cite{Zhou} was found to show unusual behaviors, such as the absence of an observable superconducting coherence peak (a critical, decisive, and indispensable signature for a superconducting gap) and a strong $k_z$ dependence of the magnitude of the gap. This behavior reported in Ref. \cite{Zhou} is inconsistent with the physical picture of the superconducting proximity effect. 

In order to understand the superconducting proximity effect in the helical surface states in the Bi$_2$Se$_3$/Bi$_2$Sr$_2$CaCu$_2$O$_{8+\delta}$ heterostructure, it is important to carefully study the electronic structure of the Bi$_2$Se$_3$ film. In this paper, we report fabrication of delicate heterostructure samples between topological insulator (TI) Bi$_2$Se$_3$ thin film and high temperature superconductor optimally doped Bi$_2$Sr$_2$CaCu$_2$O$_{8+\delta}$ ($T_{\textrm{c}}\simeq91$ K). Using angle-resolved photoemission spectroscopy, we probe the electronic structure and the possible existence of superconducting gap on the top surface of Bi$_2$Se$_3$ thin films. Our systematic data provides clear evidence for a two-(crystalline orientation) phase coexistence, and a lack of $d$-like proximity effect in contrast to previous report \cite{Zhou}. Our Fermi surface imaging data identifies major contributors in reducing the proximity-induced gap below the 5 meV range. These results correlate with our observation of momentum space separation between the Bi$_2$Se$_3$ and Bi$_2$Sr$_2$CaCu$_2$O$_{8+\delta}$ Fermi surfaces and mismatch of crystalline symmetries in the presence of a small superconducting coherence length.  These studies not only provide critical momentum space insights into the Bi$_2$Se$_3$/Bi$_2$Sr$_2$CaCu$_2$O$_{8+\delta}$ heterostructures, but also set an upper bound on the proximity induced gap for realizing much sought out Majorana fermion condition in this system.

\begin{figure}
\centering
\includegraphics[width=8.5cm]{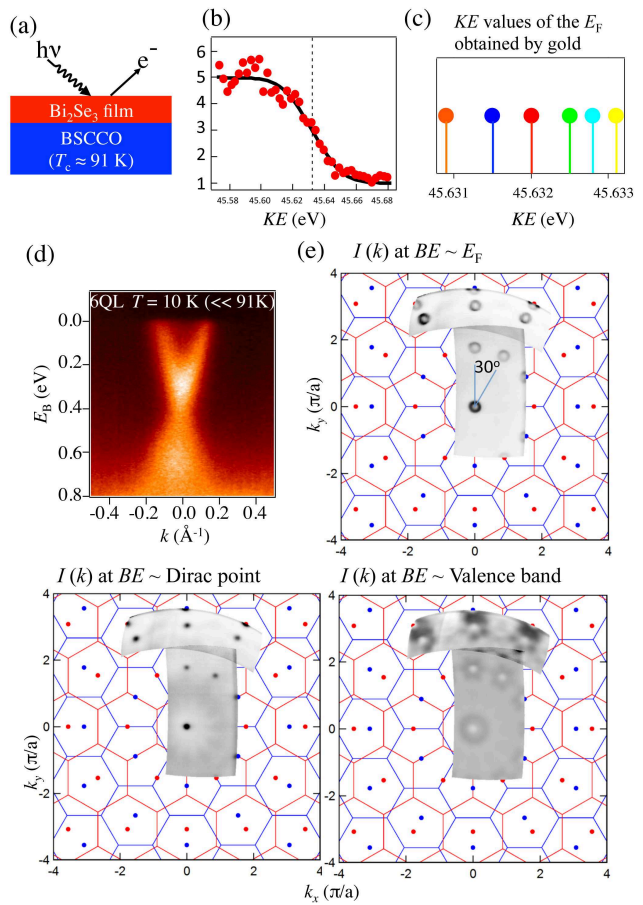}
\caption{\textbf{Characterization of Bi$_2$Se$_3$/BSCCO film heterostructure systems.} (a) Schematic illustration of our experimental configuration. (b) A gold spectrum (red circles) and its Fermi-Dirac fit (black line) measured at $T=10$ K at photon energy of 50 eV. (c) Kinetic energy values of the Fermi level determined by gold spectrum measurements at $T=10$ K at photon energy of 50 eV, at various time points during our data collection. (d) ARPES dispersion map of a 6 QL Bi$_2$Se$_3$ film sample on BSCCO. (e) ARPES Fermi surface map of a 6 QL Bi$_2$Se$_3$ film sample over a wide momentum space range superimposed on top of a schematic drawing of two sets of Bi$_2$Se$_3$ surface Brillouin zones (BZs), which are $30^{\circ}$ rotated with respect to each other. }\label{Characterization}
\end{figure}
\begin{figure*}
\centering
\includegraphics[width=17cm]{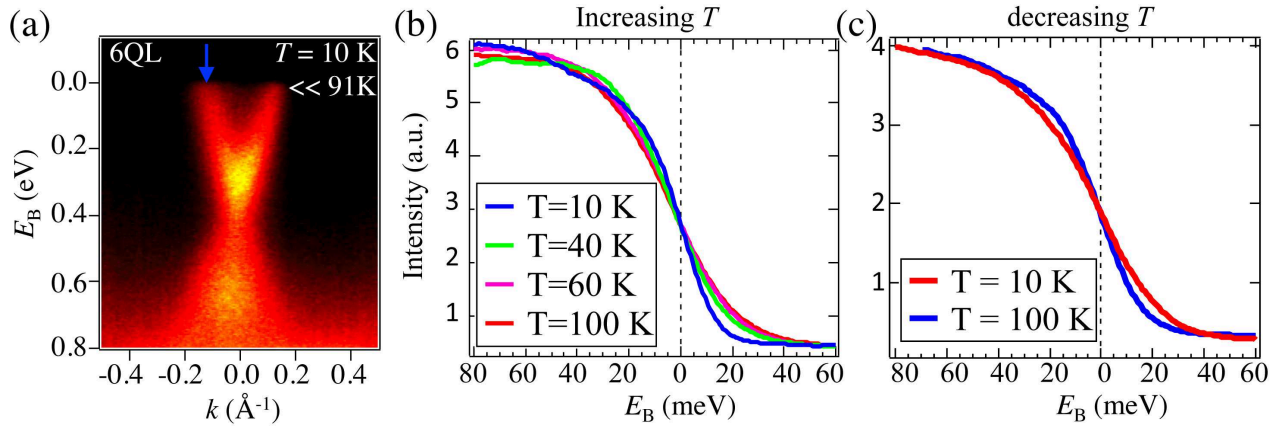}
\caption{\textbf{Temperature dependent ARPES data on a 6 QL Bi$_2$Se$_3$/BSCCO sample.} (a) ARPES dispersion map of a 6 QL Bi$_2$Se$_3$/BSCCO using photon energy of 50 eV. The blue arrow notes the momentum chosen for detailed temperature dependent studies. (b and c) ARPES energy distribution curve (EDC) data at different temperatures. For the dataset with increasing (decreasing) temperature, the measurements were taken using incident photon energy of 50 eV (55 eV). }\label{6QL}
\end{figure*}

Single crystalline samples of optimally doped Bi$_2$Sr$_2$CaCu$_2$O$_{8+\delta}$ with $T_{\textrm{c}}\simeq91$ K were grown using the standard method \cite{Gu}. The BSCCO crystals were cleaved \textit{in situ} under ultra-high vacuum, and high quality topological insulator Bi$_2$Se$_3$ thin films were then grown by the molecular beam epitaxy (MBE) on top of freshly cleaved surface of BSCCO. The MBE growth utilized thermal evaporation from high purity elemental Knudsen cells under selenium rich conditions. In order to protect the surface at the ambient pressure, a thick $\sim50$ nm selenium (Se) capping layer was deposited on the Bi$_2$Se$_3$ thin film immediately after the growth by continuing the selenium source evaporation while the film cooled to room temperature. The Se capping layer can be removed by heating the sample \textit{in situ} in the ARPES chamber at $\sim200$ $^{\circ}$C for about an hour, as reported in Ref. \cite{Hedgehog, Zhou}. High-resolution ARPES measurements were performed at the beamlines 4.0.3 and 10.0.1 at the Advanced Light Source (ALS) in the Lawrence Berkeley National Laboratory (LBNL) in Berkeley, CA. The base temperature and base pressure of the ARPES beamlines at the ALS were about 10 K and $<5\times10^{-11}$ torr, and the total energy and momentum resolution of these beamlines were about 15 meV and $0.01$ $\textrm{\AA}^{-1}$. The kinetic energy of the Fermi level was determined by fitting the ARPES spectrum of gold to the Fermi-Dirac distribution function at 10 K convolved with a gaussian function (Fig. 1b). The existence of SC gap is determined by comparing the leading edge energy shift between the energy distribution curve (EDC) in the data and the EDC in the gold spectrum. Therefore, the stability of the kinetic energy of the Fermi level over time defines the ability and stability of measuring leading-edge energies (therefore the superconducting gap) \cite{note}. We have checked the stability of the kinetic energy of the Fermi level by measuring the gold spectra over time, and have found the fluctuation of the Fermi level to be less than 2 meV in the normal running mode of the ARPES machines (Fig. 1c, and see raw data in the Supplementary Materials), at which our data on Bi$_2$Se$_3$/BSCCO films were collected. From that, we conservatively set the upper bound of SC gap at a level less than 5 meV (see Supplementary Materials for a statistical argument in setting the upper bound).

\begin{figure}
\centering
\includegraphics[width=8.5cm]{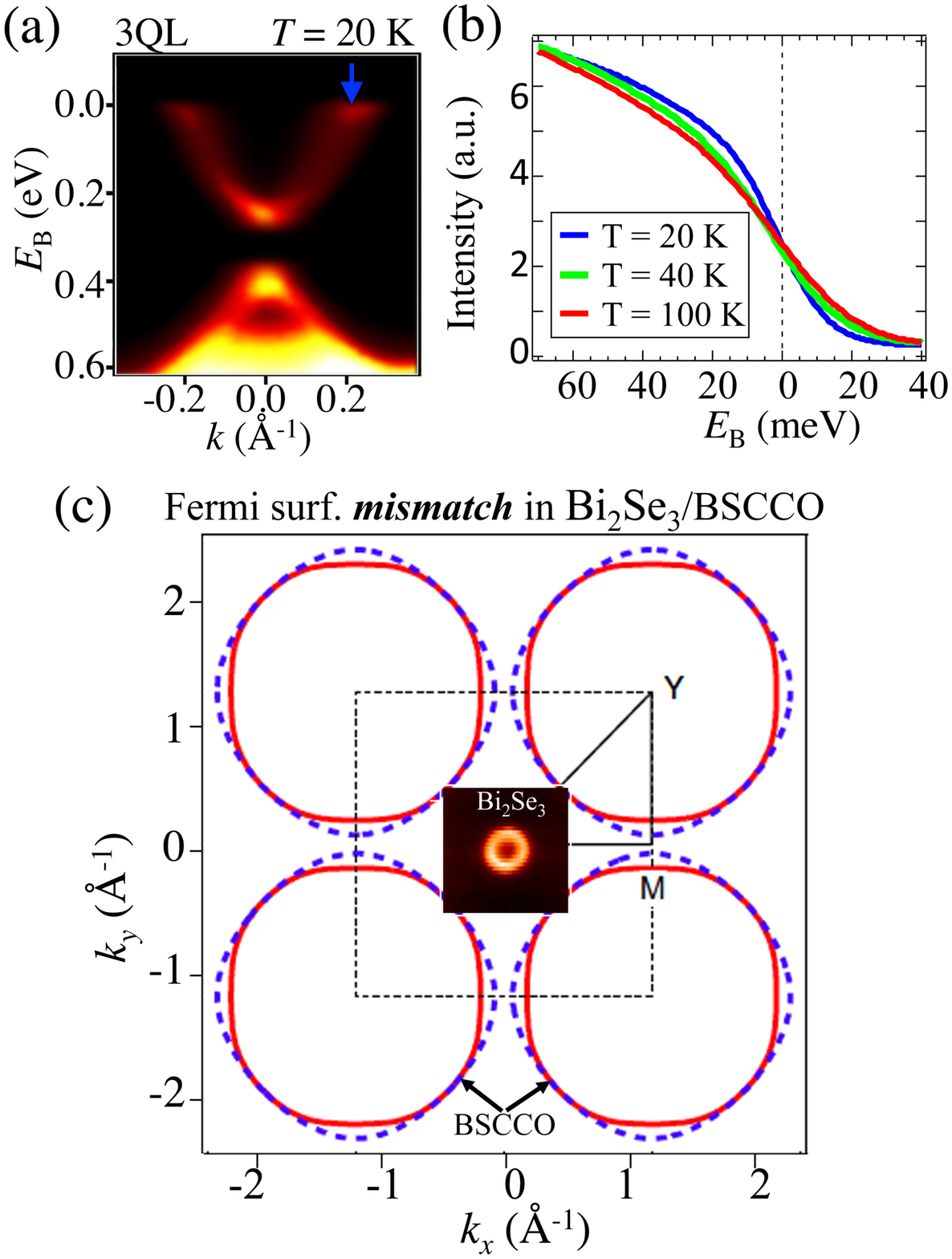}
\caption{\textbf{Temperature dependent ARPES data on a 3 QL Bi$_2$Se$_3$/BSCCO sample.} (a) Dispersion map of a 3 QL Bi$_2$Se$_3$/BSCCO sample usingphoton energy of 60 eV. The blue arrow notes the momentum chosen for detailed temperature dependent studies. (b) ARPES energy distribution curve (EDC) data ($h\nu=70$ eV) at different temperatures. (c) ARPES Fermi surface mapping of a 6 QL Bi$_2$Se$_3$ film sample superimposed on top of a schematic BSCCO Fermi surface. The two Fermi surfaces are shown using the same momentum-axes. The BSCCO Fermi surface schematic is adapted from Ref. \cite{Fink}.}\label{3QL}
\end{figure}

Fig. 1a shows the experimental configuration. Topological insulator Bi$_2$Se$_3$ thin films at various thicknesses were grown on top of freshly cleaved surface of BSCCO crystals. ARPES experiments were then performed to measure the electronic structure on the top surface of the Bi$_2$Se$_3$ films. Fig. 1d shows the energy-momentum dispersion of a 6 quintuple layer (QL) thick Bi$_2$Se$_3$ film sample on the BSCCO. A single-Dirac cone surface state centered at the $\bar{\Gamma}$ point is observed. No Dirac point gap is seen, which indicates that the 6 QL film is above the limit at which the two surfaces couple to each other, consistent with the previous report \cite{Xue Nature physics QL, Spin QL}. Fermi surface mapping over a wide momentum-space window is shown in Fig. 1e. Interestingly, the two nearby second BZ Fermi surfaces are found to be only $30^{\circ}$ rotated with respect to each other, which demonstrates that there exist two sets of BZs that are $30^{\circ}$ rotated. In real space, this observation means that the film contains two sets of crystalline (phases) domains that are $30^{\circ}$ rotated. We note that the ARPES Fermi surface in Fig. 1e  is reproducible and always contains these two sets of domains, as the beamspot is moved at different positions on a film surface. Thus the size of the domains must be intrinsically smaller than that of the beamspot ($50$ $\mu$m $\times$ $100$ $\mu$m). The coexistence of two sets of domains is a reasonable consequence considering the different lattice symmetries (square vs hexagonal) in the BSCCO substrate and the Bi$_2$Se$_3$ film.

We now study the low energy electronic structure at various temperatures across the $T_{\textrm{c}}$ of BSCCO, in order to search for possible existence of superconducting gap in the topological surface states. The blue arrow in Fig. 2a denotes the momentum where the topological surface states cross the Fermi level. The ARPES EDC data at the momentum indicated by the blue arrow at various temperatures are shown in Fig. 2b. No leading-edge shift (energy gap at the Fermi level) nor superconducting coherence peak is observed as temperature is raised from 10 K (below the $T_{\textrm{c}}$ of BSCCO) to 100 K (above the $T_{\textrm{c}}$ of BSCCO). To exclude any systematic error or artifacts, the sample is re-cooled down from 100 K to 10 K. However, again no superconducting gap is observed (Fig. 2c).

Since the 6 QL film is above the surface-to-surface coupling thickness threshold, we also study a 3 QL Bi$_2$Se$_3$ sample shown in Fig. 3. Fig. 3a shows the surface state dispersion. A gap at the Dirac point is clearly observed in the 3 QL sample, which shows that at 3 QL the top and bottom surface states are coupled to each other. Temperature dependent ARPES measurements are done at the momentum where surface states cross the Fermi level. As shown in Fig. 3b, no superconducting gap is found, which demonstrates that even at 3 QL no observable superconducting gap larger than 5 meV exists in the surface states localized near the top surface. We note that our experiments are performed with similar or better conditions than that of reported in Ref. \cite{Zhou}.


Our ARPES data exclude the existence of superconducting gap in the helical surface states of Bi$_2$Se$_3$ larger than 5 meV. Therefore, further ultra-high resolution and ultra-low temperature ARPES measurements are required to resolve the existence of a small ($\leq5$ meV) or even sub-meV superconducting gap exists in the surface states and its magnitude.  We identify the following contributors based on our data in reducing the proximity-induced superconducting gap below the $5$ meV range. First, our observation of the coexistence of two-crystalline-domains (phases) rotated by $30$ degrees demonstrate the fact that the strong mismatch of lattice symmetries between Bi$_2$Se$_3$ and BSCCO limits the quality of the interface, which is unfavorable for the large amplitude Cooper pair tunneling at the heterostructure interface, severely limiting the magnitude of proximity effect. Second, as seen in Fig. 3c, BSCCO has four pieces of Fermi surface at the Brillouin zone (BZ) corner whereas Bi$_2$Se$_3$ features one small surface Fermi surface at the surface BZ center. Therefore, the lack of momentum space overlap between their Fermi surfaces also make it difficult for the Cooper pairs tunneling across the interface. Third, the nodal $d$-wave superconducting order parameter in BSCCO is different from the theoretically expected $p+ip$ wave (isotropic SC gap, nodeless) superconductivity in TIs \cite{Kane_Proximity}. The different pairing symmetries and (nodal/nodeless) nature of the SC gap make the TI/BSCCO interface further unsuitable for a strong superconducting proximity effect. Finally, BSCCO and other cuprate superconductors are known to have a short superconducting coherence length, especially along the out-of-plane direction (only less than $1$ nm) \cite{Heine}. Thus even if the above conditions (lattice symmetries, $k_{\textrm{F}}$, superconducting order parameters and pairing symmetries) between these two systems could be perfectly matched, the proximity induced superconductivity is not expected to propagate over a distance (film thickness larger than 1 or 2 QL) along the $c$-axis of BSCCO. In addition to these factors, the $d$-wave superconductivity in BSCCO can further destabilize the Majorana fermions at the interface, since Majorana modes can leak to the BSCCO substrate through the nodes (gapless) of the superconducting gap, causing dephasing or decoherence of the zero-bias modes. Therefore, Our finding establishes a stringent criterion on proximity-induced SC gap by high temperature superconductor, indicating that it is difficult, if possible, to realize Majorana fermion in the $d$-wave proximity settings.

Apart from absence of observable superconducting gap in the surface states, the observed two micro-crystalline-domains that are $30^{\circ}$ rotated in Bi$_2$Se$_3$ worths further investigation using probes with space resolution such as STM. Furthermore, the coexistence of two domains also means that in a space-average probe such as ARPES, the $\bar{\Gamma}-\bar{M}$ and $\bar{\Gamma}-\bar{K}$ directions are now equivalent because these two directions from the two domains are mapped onto each other (Fig. 1e). Therefore, any possible anisotropy ($p_x+ip_y$) in the superconducting gap is further smeared out in ARPES measurements. Finally, it would be particularly exciting to study the proximity effect of the pseudo-gap states in the under-doped BSCCO system by searching for pseudo-gap in the Bi$_2$Se$_3$ surface states in energy scale of small sub 5 meV. 

Acknowledgement: The work at Princeton was supported by Office of Basic Energy Science, US Department
of Energy (DOE grant DE-FG-02-05ER46200). The MBE synthesis at Penn State University was supported by the
ARO MURI program. We gratefully thank G. Gu and I. Eisaki for sharing their BSCCO samples and Fang Chen for useful discussion.



\begin{thebibliography}{99}
\bibitem{Kane_Proximity} L. Fu, and C. L. Kane, \textit{Phys. Rev. Lett.} $\mathbf{100}$, 096407 (2008).
\bibitem{RMP} M. Z. Hasan, and C. L. Kane, \textit{Rev. Mod. Phys.} $\mathbf{82}$, 3045-3067 (2010).
\bibitem{Zhang_RMP} X. -L. Qi, and S. -C. Zhang, \textit{Rev. Mod. Phys.} $\mathbf{83}$, 1057-1110 (2011).
\bibitem{Zhang_TSC} X.-L. Qi, T. L. Hughes, S. Raghu, and S.-C. Zhang. \textit{Phys. Rev. Lett.} $\mathbf{102}$, 187001 (2009).
\bibitem{Linder} J. Linder, Y. Tanaka, T. Yokoyama, A. Sudb\o, and N. Nagaosa, \textit{Phys. Rev. Lett.} $\mathbf{104}$, 067001 (2010).
\bibitem{Patrick Lee} A. C. Potter, and P. A. Lee, \textit{Phys. Rev. B} $\mathbf{83}$, 184520 (2011).
\bibitem{Sarma} J. D. Sau \textit{et al}., \textit{Phys. Rev. Lett.} $\mathbf{104}$, 040502 (2010).
\bibitem{SUSY1} T. Grover, and A. Vishwanath, \textit{arXiv:1206:1332} (2012).


\bibitem{Hor} Y. S. Hor \textit{et al}., \textit{Phys. Rev. Lett.} $\mathbf{104}$, 057001 (2010).
\bibitem{Ando} S. Sasaki \textit{et al}., \textit{Phys. Rev. Lett.} $\mathbf{107}$, 217001 (2011).
\bibitem{Nitin} D. Zhang \textit{et al}., \textit{Phys. Rev. B} $\mathbf{84}$, 165120 (2011).
\bibitem{Kanigel} G. Koren \textit{et al}., \textit{Phys. Rev. B} $\mathbf{84}$, 224521 (2011).
\bibitem{Morpurgo} B. Sac\'ep\'e \textit{et al}., \textit{Nature Comm.} 2, 575 (2011).

\bibitem{LuLi1} F. Qu \textit{et al}., \textit{Scientific Reports} $\mathbf{2}$, 339 (2012).
\bibitem{Mason} S. Cho \textit{et al}.,  \textit{Nature Comm.} $\mathbf{4}$, 1689 (2013).
\bibitem{LuLi2} F. Yang \textit{et al}., \textit{Phys. Rev. B} $\mathbf{86}$, 134504 (2012).
\bibitem{Gordon} J. R. Williams \textit{et al}., \textit{Phys. Rev. Lett.} $\mathbf{109}$, 056803 (2012).

\bibitem{Dong} M.-X. Wang \textit{et al}., \textit{Science} $\mathbf{336}$, 52-55 (2012).
\bibitem{Brinkman} M. Veldhorst \textit{et al}., \textit{Nature Mater.} $\mathbf{11}$, 417 (2012).
\bibitem{Burch} P. Zareapour \textit{et al}., \textit{Nature Comm.} 3, 1056 (2012).
\bibitem{Molenkamp} L. Maier \textit{et al}., \textit{Phys. Rev. Lett.} $\mathbf{109}$, 186806 (2012).
\bibitem{Kirzhner} G. Koren, and T. Kirzhner, \textit{Phys. Rev. B} $\mathbf{86}$, 144508 (2012).
\bibitem{Millo} G. Koren \textit{et al}., \textit{Euro. Phys. Lett.}, $\mathbf{103}$ (2013) 67010.

\bibitem{LuLi3} J. Shen \textit{et al}., \textit{arXiv:1303.5598} (2013).

\bibitem{Leo} V. Mourik \textit{et al}., \textit{Science} $\mathbf{336}$, 1003-1007 (2012).
\bibitem{Shtrikman} A. Das \textit{et al}., \textit{Nature Phys.} $\mathbf{8}$, 887-895 (2012).

\bibitem{Patrick Lee2} J. Liu \textit{et al}., \textit{Phys. Rev. Lett.} $\mathbf{109}$, 267002 (2012).
\bibitem{TeWari} D. Roy, N. Bondyopadhaya, and S. Tewari, \textit{Phys. Rev. B} $\mathbf{88}$, 020502(R) (2013).
\bibitem{Marcus} H. O. H. Churchill \textit{et al}., \textit{Phys. Rev. B} $\mathbf{87}$, 241401(R) (2013)
\bibitem{Franceschi} E. J. H. Lee \textit{et al}., \textit{Nature nanotech.} $\mathbf{9}$, 79-84 (2014).

\bibitem{Zhou} E. Wang \textit{et al}., \textit{Nature physics} $\mathbf{9}$, 621-625 (2013).   


\bibitem{Gu} G. D. Gu \textit{et al}., \textit{J. Cryst. Growth} $\mathbf{137}$, 472478 (1994).

\bibitem{Hedgehog} S.-Y. Xu \textit{et al}., \textit{Nature Phys.} $\mathbf{8}$, 616-622 (2012).

\bibitem{note} We note that it is the stability of the Fermi level over time, not the energy resolution of the ARPES machine, that defines the capability and stability of measuring the superconducting gap. Finite energy resolution can Gaussian-broaden the data feature, but it is not expected to shift the energy position of the leading edge. Thus the $15$ meV energy resolution is not directly related to the capability of measuring superconducting gap. The fluctuation of Fermi level over time (tracked by measuring gold spectra over time) is found to be maximally $2$ meV. From that, we can conservatively set $5$ meV upper bound for the SC gap. See further elaboration in the Supplementary Materials.

\bibitem{Xue Nature physics QL} Y. Zhang \textit{et al}., \textit{Nature Phys.} $\mathbf{6}$, 584-588 (2010).

\bibitem{Spin QL} M. Neupane \textit{et al}., Preprint at http://arXiv:1307.5485
\bibitem{Heine} W. Lang \textit{et al}., \textit{Phys. Rev. B} $\mathbf{51}$, 9180-9192 (1995).


\bibitem{Fink} A. A. Kordyuk \textit{et al}., \textit{Phys. Rev. B} $\mathbf{67}$, 064504 (2003).

\end{thebibliography}
\end{document}